\date{}
\documentclass[aps,pra,showpacs]{revtex4-1}
\usepackage{graphicx,psfrag,amsmath,amssymb,amsfonts,latexsym,color,dcolumn}

\begin{document}

\title{Derivative expansion of the electromagnetic  Casimir energy for two
thin mirrors}

\author{C\'esar D. Fosco$^{1,2}$}
\author{Fernando C. Lombardo$^3$}
\author{Francisco D. Mazzitelli$^{1,3}$}

\affiliation{$^1$ Centro At\'omico Bariloche,
Comisi\'on Nacional de Energ\'\i a At\'omica,
R8402AGP Bariloche, Argentina}
\affiliation{$^2$ Instituto Balseiro,
Universidad Nacional de Cuyo,
R8402AGP Bariloche, Argentina}
\affiliation{$^3$ Departamento de F\'\i sica {\it Juan Jos\'e
 Giambiagi}, FCEyN UBA, Facultad de Ciencias Exactas y Naturales,
 Ciudad Universitaria, Pabell\' on I, 1428 Buenos Aires, Argentina - IFIBA}

\date{today}

\begin{abstract} 
	We extend our previous work on a derivative expansion for the
	Casimir energy, to the case of the electromagnetic field coupled to
	two thin, imperfect mirrors. The latter are described by means of
	vacuum polarization tensors localized on the mirrors.  We apply the
	results so obtained to compute the first correction to the
	proximity force approximation  to the static Casimir effect.
\end{abstract}
\pacs{12.20.Ds, 03.70.+k, 11.10.␣z}
\maketitle
\section{Introduction}\label{sec:intro}

The Casimir force is known to depend on the electromagnetic properties of
the relevant objects (`mirrors') and on their geometric configuration, in a
rather involved way~\cite{reviews}.  

To put the problem we shall deal with in context, let us consider the
Casimir force for a quite general situation, namely, we assume that the
geometry of the problem may be characterized by just two surfaces. Those
surfaces  may correspond, for example, to the boundaries of two
mirrors. Alternatively, the surfaces themselves may describe zero-width
(`thin') mirrors. Yet another possibility is that those surfaces may be 
the interfaces between media with different electromagnetic properties,
occupying different spatial regions. 
In a situation like the ones above, one can think of the Casimir energy as
a {\em functional\/} of the functions determining the surfaces.  Of course,
it is generally quite difficult to compute that functional for arbitrary
surfaces; exact results are available only for highly symmetric
configurations, the simplest of which is perhaps the case of two flat,
infinite, parallel plates. 

However, when the surfaces are gently curved, almost parallel, and close to
each other, the proximity force approximation (PFA) is expected to be a
very accurate method to calculate the Casimir energy. Introduced by
Derjaguin many years ago~\cite{Derjaguin} to compute Van der Waals forces,
this approximation consists of replacing both surfaces by a set of parallel
plates. Then one calculates the energy as the sum of the Casimir energies
due to each pair of plates (each plate paired with the nearest one in the other
mirror). 
The PFA has also been used successfully applied in other contexts, like nuclear
physics~\cite{nuclear} and electrostatics~\cite{electrostatics}.

In spite of the simplicity and long standing usefulness of the PFA, its validity had not
been possible to asses until quite recently, mostly because there was no
systematic way of improving the approximation. Indeed, even the next to
leading order (NTLO) correction was unknown.  
In a recent work~\cite{pfa_nos}, we have shown that the PFA can
be thought of as an expansion of the Casimir energy in derivatives of the
functions that describe the shapes of the surfaces. The leading order in
this expansion, that contains no derivatives, does reproduce the PFA,
while the higher order terms contain the corrections.  In Ref.~\cite{pfa_nos} we
considered the case of a flat surface in front of a gently curved one, the
latter described by a function $x_3=\psi(x_1,x_2)$. For simplicity, we computed
the vacuum energy for a massless quantum scalar field satisfying Dirichlet
boundary conditions, the result being:
\begin{equation}
E_{\rm DE} \simeq - \frac{\pi^2}{1440}\int d^2{\mathbf x_ \parallel} \;
\frac{1}{\psi^3}\left[\beta_1+\beta_2(\partial_\alpha\psi)^2\right]\, ,
\label{DE}\end{equation}
with $\beta_1=1$ and $\beta_2=2/3$. 
The first term in this expression is the PFA, while the second term is the
NTLO correction. This result has been generalized by Bimonte et al to the
case of two curved, perfectly conducting surfaces, for scalar fields
satisfying Dirichlet or Neumann boundary conditions, and also to the
electromagnetic case~\cite{pfa_mit1}. 
The results for the latter are, $\beta_1=2$ and $\beta_2=4/3(1-15/\pi^2)$.

As a validity check, it has been shown that, whenever analytic results are
available for particular geometries, the corresponding derivative expansion
correctly reproduces both the PFA and its NTLO
correction~\cite{pfa_nos,pfa_mit1}.  Moreover, initial
discrepancies~\cite{pfa_mit1} between the improved PFA and the analytic
calculations for the particular case of a cylinder in front of a
plane~\cite{bordag_cp}, has been resolved in favor of the improved PFA
after a revision of the rather involved analytic calculation for this
particular geometry~\cite{bordagteo}. 

Bimonte et al  also considered the case in which the surfaces are
interfaces between different media, with frequency-dependent permittivity~\cite{pfa_mit2}. 
In this case, the numerical coefficients $\beta_1$ and $\beta_2$ become
rather complicated functions of $\psi$ and of the dimensional constants that describe the
electromagnetic properties of the media.  

In this paper, we will extend the improved proximity force approximation to
the case of two imperfect thin mirrors. This kind of configuration
have already been considered in several previous works; for instance,  in
order to describe the interactions of plasma sheets, graphene sheets or,  more
generally, arbitrary  semi-transparent mirrors, both for the static and
dynamical Casimir effects~\cite{varios1,bordaggraph,thindce}. 

In some derivations of the Casimir energy for perfect and imperfect
mirrors, the boundary conditions at the interfaces are represented in terms
of auxiliary scalar fields coupled to the TE and TM modes of the
electromagnetic field~\cite{golest}. We will follow here a similar
approach, but developing a new formalism, based on vector auxiliary fields
that couple to the dual of the  Maxwell tensor $F_{\alpha\beta}$ evaluated
on the surfaces.  In this formalism gauge invariance is more apparent at
the different stages of the calculation. Moreover, the formalism could be
useful to address problems with more complex geometries, where it could not
be possible to describe the electromagnetic field in terms of independent
TE and TM modes. 

The paper is organized as follows. In the next Section we describe the
model and introduce the necessary definitions and conventions. In Section 3
we derive the formal expression of the vacuum energy for the
electromagnetic field, using the above mentioned formalism based on a
vector auxiliary field.  The derivative expansion for the electromagnetic
vacuum energy is presented in Section 4. We discuss the results in Section
5, where we analyze the two limiting cases of perfectly conducting and near
transparent mirrors. In the latter, we find that the NTLO correction to the
PFA is tantamount to use the area of the curved surface in the leading
order expression.  We also discuss in that section  a particular class of
imperfect mirrors, in which the transmission and reflection coefficients do
not involve dimensionful constants. For these `graphene-like' mirrors,
dimensional analysis implies that the vacuum energy is of the form given in
Eq.(\ref{DE}), where $\beta_1$ and $\beta_2$ are constants and depend on the
dimensionless quantities that describe the mirrors.  We present some
numerical evaluations of these coefficients that interpolate between almost
transparent and perfectly conducting mirrors. 
Section 6 contains the conclusions of this work. 

\section{The model: definitions and conventions}
We shall consider a model in which the role of the fluctuating vacuum field is
played by an Abelian gauge field, $A_\mu$, in $3+1$ dimensions, coupled to
two imperfect mirrors, $L$ and $R$. These are presumed to have negligible
widths, so that we shall use an idealized description whereby they are
treated as mathematical surfaces. We will, moreover, assume that one of the
surfaces ($L$) is a plane, while the other ($R$), which may be curved, can
always be described by a Monge patch. Summarizing, the two surfaces
correspond to:
\begin{equation}\label{eq:deflr}
L) \;\; x_3 \,=\, 0 \,\;\;\;\;\;\; R) \;\; x_3 \,=\, \psi(x_1,x_2) \;,
\end{equation}
where $x_i$ \mbox{($i=1,2,3$)} are the spatial coordinates.  Throughout
this work we shall use Euclidean conventions, with $x^\mu \equiv x_\mu$
\mbox{($\mu=0,1,2,3$)}, $x_0$ being the imaginary time. 
However, we have found it simpler to keep our treatment quite general
regarding the actual form of the surfaces, postponing the use of
(\ref{eq:deflr}) to the point when we actually need those particular expressions.

The action ${\mathcal S}$ for this model will have the following structure:
\begin{eqnarray}\label{eq:defs}
	{\mathcal S} &=& {\mathcal S}(A; y_L, y_R) \nonumber\\
	             &=& {\mathcal S}_0(A) \,+\, {\mathcal S}_L(A; y_L) 
		     \,+\, {\mathcal S}_R(A; y_R) \;,
\end{eqnarray}
where $A$ denotes  the electromagnetic  field and
${\mathcal S}_0$ its free action.  ${\mathcal S}_L$ and ${\mathcal
S}_R$ are terms that couple $A$ to each mirror, with  $y_L$ and $y_R$
denoting parametrizations of their respective surfaces.  ${\mathcal S}_L$ and ${\mathcal S}_R$ 
can  be different because of two
reasons: first, they correspond to different surfaces, and second, they may
also have to account for different electromagnetic (response) properties,
for example, when the mirrors are composed of different materials.

We shall then consider a rather general term, ${\mathcal S}_\Sigma$,
corresponding to the coupling to an arbitrary surface $\Sigma$,
particularizing to the $L$ and $R$ cases afterwards.  Thus, we assume the
static surface $\Sigma$ to be defined in parametric form:
\begin{equation}\label{eq:defsigma}
	\Sigma \big) \;\;\; (\sigma^1, \sigma^2) \; \to \; {\mathbf
	y}(\sigma^1, \sigma^2) \;\in\; {\mathbb R}^3\;.
\end{equation}
Although the surface is static, to write the ($2+1$-dimensional) term
${\mathcal S}_\Sigma$ it is, however, convenient, to introduce a
parametrization for the world-volume ${\mathcal V}$ swept by the surface
$\Sigma$, since that is the spacetime region ${\mathcal V}$ where the
interaction takes place:
\begin{equation}\label{eq:defcalv}
	{\mathcal V} \big) \;\;\; (\sigma^0, \sigma^1, \sigma^2) \; \to \;
	y^\mu(\sigma^0, \sigma^1, \sigma^2) \, \equiv \, y^\mu(\sigma),
	\;\;\;\;\mu=0,\,1,\,2,\,3,
\end{equation} 
where $y^0 = \sigma^0$, and  ${\mathbf y}$ as in (\ref{eq:defsigma}).
The world-volume is three-dimensional, and we adopt the convention of
using indices from the beginning of the Greek alphabet to denote components
in that space; for example, in an expression like $d\sigma^\alpha$ we
implicitly assume that $\alpha$ runs from $0$ to $2$. We do need to introduce
more objects in that space, like the induced metric, $g_{\alpha\beta}(\sigma)$,
which may be written in terms of the parametrization:
\begin{equation}\label{eq:defgab}
g_{\alpha\beta}(\sigma) = \frac{\partial y_\mu(\sigma)}{\partial \sigma^\alpha}
\frac{\partial y_\mu(\sigma)}{\partial \sigma^\beta} \;.
\end{equation}
We also need to introduce $e^\mu_\alpha$, a local basis of tangent vectors
to ${\mathcal V}$,  such that \mbox{$e^\mu_\alpha = \frac{\partial
y^\mu}{\partial \sigma^\alpha}$}. They are, by construction,
normalized to satisfy the condition \mbox{$e^\mu_\alpha(\sigma) e^\mu_\beta(\sigma) =
g_{\alpha\beta}(\sigma)$}. 

Before writing the explicit expression for the action, and to make contact with previous works,
let us describe a simpler model with a vacuum scalar field. 
The free action is 
\begin{equation}
S_0(\varphi) = \int d^4x \frac{1}{2}\partial_\mu\varphi\partial_\mu\varphi\, .
\end{equation}
Assuming that the surface action ${\mathcal S}_\Sigma$  is quadratic in $\varphi$,
its general form is 
\begin{equation}
	{\mathcal S}_\Sigma (\varphi; y) = \frac{1}{2} \int d^3\sigma \,
	\sqrt{g(\sigma)}   \, d^3\sigma' \, \sqrt{g(\sigma')}    \,
	\varphi(\sigma) \varphi(\sigma')
	{\pi(\sigma,\sigma')} \;, 
\end{equation}
where $g(\sigma)$ is the determinant of the induced metric and $\pi(\sigma,\sigma')$
describes the (nonlocal) response of the mirror. The local approximation of this action is
\begin{equation}
	{\mathcal S}_\Sigma (\varphi; y) = \frac{\lambda_s}{2}  \int
	d^3\sigma \, \sqrt{g(\sigma)}   \varphi(\sigma)^2 \;, 
	\label{deltapot}
\end{equation}
where $\lambda_s$ is a constant (the subindex $s$ stands for scalar).
Eq.(\ref{deltapot}) describes the so called `$\delta$-potentials', widely used
as toy models to describe imperfect mirrors. One can
check that this kind of potentials induce a discontinuity in the normal
derivative of the scalar field across the surface, i.e.  ${\rm disc}
[\partial_n\varphi] = \lambda_s\varphi$. The factor $ \sqrt{g(\sigma)}$ is
crucial to produce such boundary condition \cite{sqrtg}.  In the limit
$\lambda_s\to\infty$, the field must vanish on the surface in order to have
a finite discontinuity across the surface, and therefore one recovers the
usual Dirichlet boundary condition on $\Sigma$.  

In the electromagnetic case, the explicit form of ${\mathcal S}_0(A)$ will be
\begin{equation}
	S_0(A) \;=\; \int d^4x \big[ \frac{1}{4} F_{\mu\nu} F_{\mu\nu}  
	\,+\, \frac{b}{2} (\partial_\mu A_\mu)^2 \big] \;,
\end{equation}
where the term proportional to $b$ provides the gauge-fixing. In the
calculations presented in the next sections,  we shall adopt the Feynman
($b=1$) gauge.

Assuming that ${\mathcal S}_\Sigma$ is quadratic in $A_\mu$, gauge invariance
implies that it will have the general form
\begin{equation}\label{eq:gralsigma}
	{\mathcal S}_\Sigma (A; y) = \frac{1}{4} \, \int
	d^3\sigma \, \sqrt{g(\sigma)}  \, d^3\sigma' \, \sqrt{g(\sigma')}
	\, F_{\alpha\beta}(\sigma) F_{\alpha'\beta'}(\sigma')
	\pi^{\alpha\beta\alpha'\beta'}(\sigma,\sigma') \;, 
\end{equation}
where $\pi^{\alpha\beta\alpha'\beta'}$ is a polarization tensor that
depends on the microscopic degrees of freedom on the mirror, and
\begin{equation}
	F_{\alpha\beta}\,=\, \nabla_\alpha A_\beta -  \nabla_\beta A_\alpha
	\,=\, \partial_\alpha A_\beta -  \partial_\beta A_\alpha
	\;,
\end{equation}
where $\nabla_\alpha$ denotes the covariant derivative operator,
corresponding to the connection for the induced metric, acting
(in this case) on a covariant vector.  We have used $A_\alpha (\sigma)$ as
a shorthand for the components of the gauge field $A_\mu(x)$ on ${\mathcal
V}$, projected along the directions defined by the local basis:
\begin{equation}\label{eq:defasigma}
	A_\alpha(\sigma) \;\equiv\; A_\mu[y(\sigma)] \,
	e^\mu_\alpha(\sigma) \;.
\end{equation}

As in the scalar case, we will start our discussion with a local interaction
\begin{equation}\label{eq:defssigma}
	{\mathcal S}_\Sigma (A; y) \;=\; \frac{\lambda}{4} \, \int
	d^3\sigma \, \sqrt{g(\sigma)}\, F_{\alpha\beta} F^{\alpha\beta} \;, 
\end{equation}
where 
$\lambda$ is a constant, and afterwards  we shall
extend the results to include frequency-dependent couplings. Note that the constants
$\lambda$ and $\lambda_s$ have different dimensions.

As it should be evident from the
actual form of the interaction term that we are assuming for the model,
this kind of mirror involves only on the gauge field components
which are {\em parallel} to the world-volume. However, a term like this
induces discontinuities across the surface of
the component of the electric field which is normal to
$\Sigma$, and of the components of the magnetic field which are parallel to
that surface. The discontinuity is proportional to $\lambda$ and depends on
parallel components of the gauge field, producing the boundary conditions of
a perfect conductor in the limit $\lambda\to\infty$, as in the scalar case.
For instance, for the flat surface
at $x_3=0$ the boundary conditions read
\begin{equation}
\rm{disc}(F_{3\nu})=\lambda\, \partial_{\hat\mu}F_{\hat\mu\nu}\, ,
\label{bc}
\end{equation}
where the sum over $\hat\mu$ excludes $\mu=3$.  These boundary conditions can be 
 explicitly written in terms of the field components as
\begin{eqnarray}
\rm{disc}(E_3) &=& \lambda \partial_1 E_1,\nonumber \\
\rm{disc}(B_2) &=& -\lambda \partial_0 E_1,\nonumber\\
\rm{disc}(B_1) &=& - \lambda \left( \partial_0 E_2 + \partial_1 B_3 \right ),\nonumber
\end{eqnarray}
where, for simplicity,  we assumed that the fields do not depend on the coordinate $x_2$.

Note that in Eq.(\ref{eq:defssigma}) we are assuming a Lorentz-invariant interaction, which would be 
produced by relativistic degrees of freedom on the mirror. One could of course consider the interaction between
the gauge fields and non-relativistic matter on the mirror,  giving different boundary conditions \cite{barton2004,plb2008}.
For example, the boundary conditions obtained in the case of a fluid of nonrelativistic electrons \cite{barton2004} coincide 
with ours for the normal component of the electric field, but differ for the parallel components of the magnetic field.  
Accordingly, the reflection and transmission coefficients in both models will be different.

Finally,  one should check that a term like
(\ref{eq:defssigma}) does preserve gauge invariance.  This is indeed
the case that can be seen from the fact that under the transformation:
\mbox{$A_\mu(x) \to A_\mu(x) + \partial_\mu \omega(x)$}, which is a $U(1)$
gauge transformation in $3+1$ dimensions, one has  \mbox{$A_\alpha(\sigma)
\to A_\alpha(\sigma) + \delta A_\alpha(\sigma)$}, with:
\begin{equation}
\delta A_\alpha(\sigma) = \partial_\mu\omega[y(\sigma)] \, e^\mu_\alpha(\sigma)
 = \partial_\mu\omega[y(\sigma)]  \partial_\alpha y^\mu(\sigma)
 = \partial_\alpha \omega(\sigma)
\end{equation} 
where $\omega(\sigma) \equiv \omega[y(\sigma)]$. Thus $S_{\Sigma}$ is
invariant, since $\delta F_{\alpha\beta}(\sigma) = 0$. 

\section{Electromagnetic vacuum energy: auxiliary vector fields}

In the functional approach to the Casimir effect, to obtain the vacuum
energy, one usually starts from ${\mathcal Z}$, the vacuum transition
amplitude, or, equivalently, the zero temperature limit of a
finite-temperature partition function. For the vacuum field $A_\mu$, in the
presence of the two mirrors, the case at hand, 
${\mathcal Z}$ may be written as follows:
\begin{equation}\label{eq:dezfda} 
	{\mathcal Z}\;=\; \int {\mathcal D}A_\mu \;
	\exp\Big[- {\mathcal S}_0(A) - {\mathcal S}_L(A; y_L) - {\mathcal
S}_R(A; y_R) \Big]\;.
\end{equation}
To proceed, one should integrate the electromagnetic field. To that end,
it is convenient to perform first a transformation of the interaction
terms, so that $A_\mu$ only appears linearly, rather than quadratically.  This
may be done at the expense of introducing auxiliary fields, a procedure
that we implement now. To simplify the procedure, we first represent
$F_{\alpha\beta}$ in terms of its dual $\widetilde{F}^\alpha$, a pseudo-vector, such that:
\begin{equation}\label{eq:dual}
	F_{\alpha\beta} \,=\, \sqrt{g(\sigma)} 
	\,\epsilon_{\alpha\beta\gamma} \, \widetilde{F}^\gamma \;,\;\;\;
 \sqrt{g(\sigma)} \widetilde{F}^\alpha \,=\,\epsilon^{\alpha\beta\gamma}
\partial_\beta A_\gamma \;,
\end{equation}
where we adopted the convention that $\epsilon_{\alpha\beta\gamma}$, as
well as $\epsilon^{\alpha\beta\gamma}$, denote the Levi-Civita permutation
symbol (i.e., without including any power of $g$ as a factor).

The generic term $S_\Sigma$ may be written as 
\begin{equation}
{\mathcal S}_\Sigma (A; y) \;=\; \frac{\lambda}{2} \, \int
d^3\sigma \, \sqrt{g(\sigma)}\; \widetilde{F}_\alpha \, {\widetilde F}^\alpha \;. 
\end{equation}
Then we introduce a pseudo-vector auxiliary field $\xi_\alpha(\sigma)$, so
that the exponential of the interaction term above may be
obtained as the result of a Gaussian integral:
\begin{equation}\label{eq:exponent}
	\exp\left [-{\mathcal S}_\Sigma (A; y) \right] \;=\; \frac{1}{\mathcal N_\xi}\int {\mathcal D} \xi \; 
\exp \big[ - {\mathcal S}_q(\xi;\lambda) + i \int d^3\sigma \sqrt{g(\sigma)} \xi_\alpha(\sigma)
\widetilde{F}^\alpha(\sigma) \big] \;,
\end{equation}
where $S_q(\xi;\lambda) = \frac{1}{2 \lambda} \int d^3\sigma \sqrt{g(\sigma)} \xi_\alpha(\sigma)
\xi^\alpha(\sigma)$ and
\begin{equation}
	{\mathcal N_\xi} \;=\; \int {\mathcal D} \xi \; e^{-{\mathcal
	S}_q(\xi;\lambda)} \;.
\end{equation}

Note that the representation above is not unique, in the following sense: 
defining the longitudinal ($l$) and
transverse ($t$) components of $\xi$:
\begin{equation}
	\xi_l^\alpha(\sigma) \;=\; \nabla^\alpha \frac{1}{\Delta}
	\nabla_\beta \xi^\beta \;,\;\;\;
	\xi_t^\alpha(\sigma) \;=\; \xi^\alpha(\sigma) \,-\,
\xi_l^\alpha(\sigma)\;,
\end{equation}
we see that $\xi_l$ does not couple to $A_\alpha$.  Indeed,  because of (\ref{eq:dual}), we see that: 
\begin{equation}
\int d^3\sigma \sqrt{g(\sigma)} \xi^\alpha(\sigma)
\widetilde{F}_\alpha(\sigma) \,=\, \int d^3\sigma \sqrt{g(\sigma)} \xi_t^\alpha(\sigma)
\widetilde{F}_\alpha(\sigma)\, ,
\end{equation} 
where we have used Bianchi's identity:
\mbox{$\nabla^\alpha\widetilde{F}_\alpha=\epsilon^{\alpha\beta\gamma}\partial_\alpha \partial_\beta
A_\gamma=0$}. 

It is, therefore, possible to modify the auxiliary field action, for example
by adding a term depending only on $\xi_l$ to the ${\mathcal S}_q$ term,
such that:
\begin{equation}
{\mathcal S}_q \to {\mathcal S}'_q = {\mathcal S}_q + 
{\mathcal S}_{\xi_\Sigma}\, ,
\end{equation} 
where ${\mathcal S}_{\xi_\Sigma}$  is a function of $\xi^l$ (and  not of
$\xi_t$) whose precise form will be determined in order to simplify the
calculations.
Thus, a more general (but equally valid) way to rewrite the interaction term is 
\begin{equation}\label{eq:exponentg}
	\exp\big[-{\mathcal S}_\Sigma (A; y) \big] = \frac{1}{\mathcal N'_\xi} \, \int {\mathcal D}
\xi \; \exp \Big[- {\mathcal S'}_q(\xi;\lambda) + i \int d^3\sigma
\sqrt{g(\sigma)} \, \xi_\alpha(\sigma) \widetilde{F}^\alpha(\sigma) \Big] \;,
\end{equation}
\begin{equation}
	{\mathcal N'_\xi} \;\equiv\; \int {\mathcal D} \xi \; e^{-{\mathcal
	S'}_q(\xi;\lambda)} \;.
\end{equation}

Besides, note that the term which couples linearly the gauge field to the
auxiliary field, can be reinterpreted as an interaction with a
surface-dependent `current' $J^\mu_\Sigma(x)$: 
\begin{equation}
\int d^3\sigma \xi_\alpha(\sigma)
\epsilon^{\alpha\beta\gamma}\partial_\beta A_\gamma(\sigma) \,=\, \int d^4x
J^\mu_\Sigma(x) \, A_\mu(x)\;,
\end{equation}
where
\begin{equation}\label{eq:jsigma}
J^\mu_\Sigma(x) \;=\; \int d^3\sigma \, \delta^{(4)}[x-y(\sigma)] \,
e^\mu_\alpha(\sigma) \, \epsilon^{\alpha\beta\gamma}\partial_\beta
\xi_\gamma(\sigma) \;,
\end{equation}
is a `topologically conserved' current, namely, it satisfies
\mbox{$\partial_\mu J^\mu_\Sigma = 0$}, by its very form, regardless of dynamics.

The process introduced above for $\Sigma$ may be then independently applied to the two
interaction terms, ${\mathcal S}_L$ and ${\mathcal S}_R$, which are
defined as follows: 
\begin{eqnarray}
	{\mathcal S}_L(A;y_L) &=& {\mathcal S}_\Sigma (A; y)
	\Big|_{\Sigma \to L,\, y \to y_L, \, \lambda \to \lambda_L} \nonumber\\
	{\mathcal S}_R(A;y_R) &=& {\mathcal S}_\Sigma (A; y) \Big|_{\Sigma
	\to R,\, y \to y_R,\,\lambda \to \lambda_R}\;.
\end{eqnarray}
Since we introduce one auxiliary field for each interaction term in the
action, the gauge field will be coupled linearly to the sum of two
currents. Indeed, the  use of two auxiliary fields, $\xi_L$ and $\xi_R$, in the
partition function, yields:
$${\mathcal Z} \;=\; \frac{1}{\mathcal N'_{\xi_L} \mathcal
N'_{\xi_R}} \int {\mathcal D}\xi_L {\mathcal D}\xi_R 
	 \; e^{- {\mathcal S'}_q(\xi_L;\lambda_L) - {\mathcal S'}_q(\xi_R;\lambda_R)}$$ 
\begin{equation}\label{eq:zetal}
\times \,\int {\mathcal D}A \, e^{-{\mathcal S}_0(A) + i  \int d^4x \,
J^\mu(x)A^\mu(x)}
\end{equation}
where $J \equiv J_L + J_R$, with $J_L$ and $J_R$ obtained from (\ref{eq:jsigma}), replacing
$\Sigma$ by $L$ and $R$, respectively.

Integrating out $A_\mu$ in (\ref{eq:zetal}), yields:
\begin{eqnarray}\label{eq:zetal1}
{\mathcal Z} \;=\; \frac{\mathcal Z_0}{\mathcal N'_{\xi_L} \mathcal
N'_{\xi_R}}  \; \int {\mathcal D}\xi_L {\mathcal D}\xi_R 
\; \Big\{ e^{- {\mathcal S'}_q(\xi_L;\lambda_L) - {\mathcal S'}_q(\xi_R;\lambda_R)}
 \nonumber\\
\times \exp\big[- \frac{1}{2} \int d^4x \int d^4x' J^\mu(x)
D_{\mu\mu'}(x,x') J^{\mu'}(x')\big] \Big\} \;,
\end{eqnarray} 
where $D_{\mu\mu'}$ is the (free) $A$-field propagator, which in the Feynman
gauge becomes:
\begin{equation}\label{eq:propa}
D_{\mu\mu'}(x,x') \;=\; \delta_{\mu \mu'} \; D(x,x')\; .
\end{equation}
Here $D(x,x')$ is the Euclidean free scalar field propagator in $3+1$
dimensions:
\begin{equation}\label{eq:defscp}
D(x,x') \,=\, \langle x| \frac{1}{-\partial^2} |x'\rangle \,=\,
\int \frac{d^4k}{(2\pi)^4} \, \frac{e^{i k \cdot (x-x')}}{k^2} \, ,
\end{equation}
where we have used a `bra-ket' notation to denote matrix elements of
functional operators. Note that ${\mathcal Z}_0 = \int {\mathcal D}A \,
e^{-{\mathcal S}_0}$ cannot contribute to the
Casimir energy, since it is independent of the coupling to the mirrors. On
the other hand, the normalization factors ${\mathcal N'_{\xi_L}}$
${\mathcal N'_{\xi_R}}$ do not contribute either, albeit for a different
reason: each one of them depends only on the properties of one the mirrors,
being adamant to the coupling of the other.  Thus, we define the vacuum
energy, $E_{\rm vac}$, in such a way that those contributions, irrelevant to
the Casimir interaction energy, are subtracted from the very beginning: 
\begin{equation}\label{eq:defgamma}
E_{\rm vac} \,=\, \lim_{T \to \infty} \Big( \frac{\Gamma}{T} \Big) \;,\;\;\;\; 
e^{-\Gamma} \;\equiv\; {\mathcal Z} \;
\frac{{\mathcal N'_{\xi_L}}{\mathcal N'_{\xi_R}}}{{\mathcal Z}_0}\;, 
\end{equation}
where $T$ is the extent of the imaginary time interval.

We then proceed to the evaluation of $\Gamma$, defined in
(\ref{eq:defgamma}). We note that there still remain in this object
contributions that correspond to mirrors' self-interactions, depending on
only one of the mirrors. They will be neglected, since our objective is to
calculate the Casimir interaction energy between two mirrors, a physical
magnitude to which self-interaction energies cannot contribute.

We deal now with the functional integral expression for $\Gamma$,
which in view of the above has the following structure:
\begin{equation}\label{eq:gamma0}
e^{-\Gamma} \,=\, \int {\mathcal D}\xi_L {\mathcal D}\xi_R \; e^{- S_\Gamma (\xi_L, \xi_R) }
\end{equation} 
where 
\begin{eqnarray}
	S_\Gamma (\xi_L, \xi_R) &=& {\mathcal S}_q(\xi_L;\lambda_L) + {\mathcal
S}_q(\xi_R;\lambda_R) \nonumber\\
&+& \frac{1}{2} \int d^4x \int d^4x' J_L^\mu(x) D_{\mu\mu'}(x,x')
J_L^{\mu'}(x') \nonumber\\
&+&\frac{1}{2} \int d^4x \int d^4x' J_L^\mu(x) D_{\mu\mu'}(x,x')
J_R^{\mu'}(x') \nonumber\\
&+&\frac{1}{2} \int d^4x \int d^4x' J_R^\mu(x) D_{\mu\mu'}(x,x')
J_L^{\mu'}(x') \nonumber\\
&+&\frac{1}{2} \int d^4x \int d^4x' J_R^\mu(x) D_{\mu\mu'}(x,x') J_R^{\mu'}(x') \;,
\end{eqnarray}
which is a quadratic form in the auxiliary fields. In order to perform the
integral over the auxiliary fields, we need an explicit form for the
different terms in ${\mathcal S}_\Gamma$.

Taking into account (\ref{eq:deflr}), we can find the metric tensors and
local tangent vector for each mirror; all of these are elements
that enter in the terms above. In both cases, the parameters
$\sigma^\alpha$ are chosen as $\sigma^\alpha = x_\alpha$, with 
$\alpha=0,1,2$. We also refer to $(x_\alpha)$ as $x_\parallel$, reserving the
notation ${\mathbf x}_\parallel$ for $(x_1,x_2)$.

For the  $L$ surface, the parametrization is then:
\begin{equation}
x_\parallel \to y_L(x_\parallel),\;\; 
y_L(x_\parallel)=( x_\parallel, 0) \;,
\end{equation}
thus, for $L$ we simply have \mbox{$g_{\alpha\beta} =
\delta_{\alpha\beta}$}, $e^\mu_\alpha = \delta^\mu_\alpha$ for $\mu=0,1,2$,
while $e^3_\alpha=0$. For $R$, on the other hand: 
\begin{equation}
x_\parallel \to y_R(x_\parallel),\;\; 
y_R(x_\parallel)=( x_\parallel, \psi({\mathbf x}_\parallel)) \;.
\end{equation}
Therefore,
\begin{equation}
	(g_{\alpha\beta}) \;=\; \left( 
	\begin{array}{ccc} 
		1 & 0 & 0 \\
		0 & 1 + (\partial_1 \psi)^2  & \partial_1\psi \partial_2
		\psi  \\
		0 & \partial_2\psi \partial_1\psi  & 1 + (\partial_2\psi)^2   
	\end{array}
	\right) \;,
\end{equation}
which implies: $\sqrt{g} = \sqrt{ 1 + ({\nabla \psi})^2}$. The tangent
vectors, on the other hand, are given by:
\begin{equation}
e^\mu_\alpha(x_\parallel) \,=\, \delta^\mu_\alpha \,+\,
\partial_\alpha\psi({\mathbf x}_\parallel) \delta^\mu_3 \;
=\; \left\{ 
\begin{array}{ccc}
\delta^\mu_0 & {\rm if} & \alpha = 0 \\
\delta^\mu_i + \partial_i \psi({\mathbf x}_\parallel) \delta^\mu_3 & {\rm
if} & \alpha = i = 1, 2 \;.
\end{array}
\right.
\end{equation}

Then we find that:
$$
\int_{x,x'} J_L^\mu(x) D_{\mu\mu'}(x,x') J_L^{\mu'}(x') 
$$
\begin{equation}
= \int_{x_\parallel,x'_\parallel} \xi^L_\alpha(x_\parallel)
\;(\partial_\beta \partial'^\beta \delta^{\alpha\alpha'} -
\partial^{\alpha'}
{\partial'}^{\alpha})
 D(x_\parallel, 0 ; x'_\parallel,0) \; \xi^L_{\alpha'}(x'_\parallel)
\end{equation}
where we have adopted the notations: $\partial^\alpha \equiv
\partial/ {\partial x_\alpha}$, ${\partial'}^{\alpha} \equiv
\partial/{\partial x'_{\alpha}}$, etc. Besides, we have written the integration
variables as a subindex of the integral.

For the analogous term that involves the $J_R$ current instead of $J_L$,
the corresponding expression is: 
$$
\int_{x,x'} J_R^\mu(x) D_{\mu\mu'}(x,x') J_R^{\mu'}(x') 
$$
\begin{eqnarray}
= \int_{x_\parallel,x'_\parallel} \; \xi^R_\alpha(x_\parallel)
\Big\{ \big[(\partial_\beta \partial'^\beta \delta^{\alpha\alpha'} -
\partial^{\alpha'} {\partial'}^\alpha) 
D(x_\parallel, \psi({\mathbf x}_\parallel); 
x'_\parallel,\psi({\mathbf x}'_\parallel)) \big]
\nonumber\\
+ \epsilon^{\alpha\beta\gamma}
\epsilon^{{\alpha'}{\beta'}{\gamma'}}
\partial_\beta\psi(x_\parallel) \, \partial'_{\beta'}
\psi({x'}_\parallel) 
\big[\partial_\gamma \partial'_{\gamma'} 
D(x_\parallel, \psi({\mathbf x}_\parallel) ; 
x'_\parallel,\psi({\mathbf x}'_\parallel))\big] \Big\}
\xi^R_{\alpha'}(x'_\parallel) \;.
\end{eqnarray}
Finally, 
$$
\int_{x,x'} J_L^\mu(x) D_{\mu\mu'}(x,x') J_R^{\mu'}(x') 
$$
\begin{equation}
= \int_{x_\parallel,x'_\parallel} \xi^L_\alpha(x_\parallel)
\;(\partial_\beta \partial'^\beta \delta^{\alpha\alpha'} -
\partial^{\alpha'}
{\partial'}^{\alpha})
 D(x_\parallel, 0 ; x'_\parallel,\psi({\mathbf x}'_\parallel)) \; \xi^R_{\alpha'}(x'_\parallel)
\end{equation}
and
$$
\int_{x,x'} J_R^\mu(x) D_{\mu\mu'}(x,x') J_L^{\mu'}(x') 
$$
\begin{equation}
= \int_{x_\parallel,x'_\parallel} \xi^R_\alpha(x_\parallel)
\;(\partial_\beta \partial'^\beta \delta^{\alpha\alpha'} -
\partial^{\alpha'} {\partial'}^{\alpha})
D(x_\parallel, \psi({\mathbf x}_\parallel) ; x'_\parallel,0) 
\; \xi^L_{\alpha'}(x'_\parallel) \;.
\end{equation}
Defining the matrix kernel ${\mathbb T}$, such that
\begin{equation}
{\mathcal S}_\Gamma \;=\; \frac{1}{2} \int_{x_\parallel,x'_\parallel} 
\xi^a_\alpha(x_\parallel) \, {\mathbb T}_{\alpha \alpha'}^{ab}(x_\parallel, x'_\parallel)
\xi^b_{\alpha'}(x'_\parallel) \;,
\end{equation}
where $a,b = L, R$, the vacuum energy $E_{\rm vac}$ may be written as follows:
\begin{equation}
E_{\rm vac}\;=\; \lim_{T \to \infty} \Big[\frac{1}{2 T} {\rm Tr}\ln{\mathbb T} \Big] \;
\end{equation}
where the trace affects both continuum and discrete indices.

\section{The derivative expansion in the electromagnetic case}

As already stressed, the vacuum energy depends nontrivially on the shape of the $R$
surface, i.e. it can be thought as a nonlocal functional of $\psi$. When the R surface is gently
curved, almost parallel and close to the $L$-plane, we expect this functional to be 
well approximated by a derivative expansion:
 \begin{eqnarray}
E_{\rm vac} &\simeq&\,  \int d^2{\mathbf x}_\parallel\left[V_{\rm eff}(\psi)+Z(\psi)(\partial_j\psi)^2+\dots\right]\, \nonumber \\
&=&\, E_{\rm vac}^{(0)} + E_{\rm vac}^{(2)} + ...
\end{eqnarray}
In order to evaluate the functions $V_{\rm eff}$ and $Z$, it is enough to consider a class of surfaces of the
form $\psi({\mathbf x}) = a + \eta({\mathbf x})$ with $\eta\ll a$. Indeed, for 
these surfaces, and up to quadratic order in $\eta$, the vacuum energy
will be of the form
\begin{equation}
E_{\rm vac} \simeq\,  \int d^2{\mathbf x}_\parallel\left[V_{\rm eff}(a)+V_{\rm eff}'(a)\eta+Z(a)(\partial_j\eta)^2+\dots\right]\, ,
\label{withlinear}
\end{equation}
and therefore one can obtain $V_{\rm eff}$ and $Z$ from this expression (note that the term linear in $\eta$ vanishes
if $a$ is chosen to be the mean value of the distances between surfaces).
Therefore,
it is sufficient to perform an expansion
of $\Gamma$ in powers of $\eta$, keeping  terms with up to two
derivatives of $\eta$.

Denoting by $\Gamma^{(n)}$ and ${\mathbb T}^{(n)}$ the order-$n$ terms 
in the respective expansions for $\Gamma$ and ${\mathbb T}$, we see that, up
to the second order, the expansion for the former is given by:
\begin{equation}
\Gamma \;=\;\Gamma^{(0)} \,+\,\Gamma^{(1)}\,+\,\Gamma^{(2)}\,+\,\ldots
\end{equation}
where the zeroth and first order terms are:
\begin{eqnarray}
\Gamma^{(0)} &=& \frac{1}{2} {\rm Tr}\ln\big[{\mathbb T}^{(0)}\big] \nonumber\\
\Gamma^{(1)} &=& \frac{1}{2} {\rm Tr}\big[ \big({\mathbb T}^{(0)}\big)^{-1}
{\mathbb T}^{(1)} \big] \;,
\end{eqnarray}
while the second order term receives two contributions $\Gamma^{(2)} =
\Gamma^{(2,1)} \,+\,\Gamma^{(2,2)}$, where:
\begin{eqnarray}
\Gamma^{(2,1)} = &\frac{1}{2}& {\rm Tr}\big[ \big({\mathbb T}^{(0)}\big)^{-1} {\mathbb T}^{(2)} \big] 
\nonumber\\
\Gamma^{(2,2)} = - &\frac{1}{4}& {\rm Tr}\big[ 
\big({\mathbb T}^{(0)}\big)^{-1} {\mathbb T}^{(1)} 
\big({\mathbb T}^{(0)}\big)^{-1} {\mathbb T}^{(1)} \big] \;.
\end{eqnarray}

Let us now write the matrices ${\mathbb T}^{(j)}$, for $j=0,1,2$: 
\begin{equation}
{\mathbb T}^{(j)} \;=\; 
 \left( 
       \begin{array}{cc}
    {\mathbb T}_{LL}^{(j)} & {\mathbb T}_{LR}^{(j)} \\ 
    {\mathbb T}_{RL}^{(j)} & {\mathbb T}_{RR}^{(j)} 
      \end{array}
\right) \;.
\end{equation}
Those matrices are not completely defined until we adopt a specific form
for the action $S'_q$, which contains an arbitrary part that depends on
the longitudinal component of the auxiliary field $\xi_\alpha$.
In order to render the zeroth-order term as simple as possible, it is
convenient to add the following terms:
\begin{eqnarray}
S_{\xi_L} &=& \frac{1}{2} 
\int_{x_\parallel,x'_\parallel} 
\partial \cdot \xi_L(x_\parallel) \langle x_\parallel | 
\frac{1}{2\sqrt{-\partial^2}}|x'_\parallel \rangle  \partial \cdot \xi_L(x'_\parallel) 
\nonumber\\
S_{\xi_R} &=& \frac{1}{2} 
\int_{x_\parallel,x'_\parallel} \, \sqrt{g({\mathbf x}_\parallel)} \, 
\sqrt{g({\mathbf x}'_\parallel)} \; 
\nabla \cdot \xi_R(x_\parallel) \langle x_\parallel | 
\frac{1}{2\sqrt{-\nabla^2}}|x'_\parallel \rangle \nabla 
\cdot \xi_R(x'_\parallel) \;,
\end{eqnarray}
where we have used the `bra-ket' notation again, this time for a three dimensional space of coordinates. Besides, the derivations are understood also to act on functions defined on this space of coordinates. 

The ${\mathbb T}^{(0)}$ matrix elements, which are invariant under
translations along $x_\parallel$, may be Fourier transformed:
\begin{equation} 
{\mathbb T}^{(0)}(x_\parallel,x'_\parallel) = \int \frac{d^3k}{(2\pi)^3} 
e^{i k \cdot (x_\parallel-x'_\parallel)} \widetilde{\mathbb T}^{(0)}(k) \;, 
\end{equation}
and the explicit form of its matrix elements, for the gauge-fixing introduced above, is:
\begin{eqnarray}
    \big[\widetilde{\mathbb T}_{LL}^{(0)}\big]_{\alpha\alpha'}(k) &=& \Big(
\frac{1}{\lambda_L} + \frac{|k|}{2} \Big) \delta_{\alpha \alpha'}\nonumber\\ 
    \big[\widetilde{\mathbb T}_{RR}^{(0)}\big]_{\alpha\alpha'}(k) &=& 
\Big( \frac{1}{\lambda_R} + \frac{|k|}{2}  \Big) \delta_{\alpha \alpha'}\nonumber\\ 
\big[\widetilde{\mathbb T}_{LR}^{(0)}\big]_{\alpha\alpha'}(k) &=&
\big[\widetilde{\mathbb T}_{RL}^{(0)}\big]_{\alpha\alpha'}(k) \;=\;
\frac{|k|}{2} \big(\delta_{\alpha\alpha'} - \frac{k_\alpha
k_{\alpha'}}{k^2} \big) e^{- |k| a} \;,
\end{eqnarray}
with $|k|  \equiv \sqrt{k^2}$.

Regarding the terms of order $1$ in $\eta$, it is quite straightforward to
see that: 
\begin{equation}
{\mathbb T}_{LL}^{(1)} =  {\mathbb T}_{RR}^{(1)} = 0 \;, 
\end{equation}
so that, after evaluating the terms that mix $L$ and $R$, the result may be
put in the form: 
\begin{eqnarray}
\big[{\mathbb T}^{(1)}]_{\alpha {\alpha'}}(x_\parallel,x'_\parallel) &=& - \frac{1}{2} \big(
\partial_\beta \partial'_\beta \delta_{\alpha\alpha'} - \partial_{\alpha'}
\partial'_{\alpha} \big) \, \left( \begin{array}{cc}  0 & \eta({\mathbf x}_\parallel) \\ \eta({\mathbf x'}_\parallel)
& 0 \end{array}\right) \nonumber\\
&\times&  \, 
\int \frac{d^3k}{(2\pi)^3} \, 
e^{i k \cdot (x_\parallel - x'_\parallel)} \, e^{- |k| a} \;.
\end{eqnarray}
 
Finally, we consider the second order matrix elements. We shall also
discard terms involving more than two derivatives
of $\eta$. Since the Levi-Civita connection involves at least three
derivatives of $\eta$, we replace $\nabla$ by $\partial$ in the gauge
fixing term $S_{\xi_R}$. Thus, this term will contribute to the second
order matrix element $RR$ only through the factor depending on the
determinant of the metric. 

Besides, we see that ${\mathbb T}_{LL}^{(2)} \;=\; 0 $, while: 
\begin{eqnarray}\label{eq:tlr2}&&
\big[{\mathbb T}_{LR}^{(2)}\big]_{\alpha {\alpha'}}(x_\parallel,x'_\parallel) 
\;=\; \big[{\mathbb T}_{RL}^{(2)}\big]_{{\alpha'} \alpha'}(x'_\parallel,x_\parallel)
\nonumber\\
&=&\frac{1}{4} \big(
\partial_\beta \partial'_\beta \delta_{\alpha\alpha'} - \partial_{\alpha'}
\partial'_\alpha \big) 
 [\eta({\mathbf x}_\parallel)]^2 
\int \frac{d^3k}{(2\pi)^3} e^{i k \cdot (x_\parallel - x'_\parallel)} \, |k|
e^{- |k| a} \;.
\end{eqnarray}
Regarding the $RR$ matrix element, we have four different terms:
\begin{equation}
{\mathbb T}_{RR}^{(2)} \,=\, {\mathbb T}_{RR}^{(2,1)} +{\mathbb
T}_{RR}^{(2,2)}+
{\mathbb T}_{RR}^{(2,3)} + {\mathbb T}_{RR}^{(2,4)} 
\end{equation}	
where
\begin{eqnarray}\label{eq:t2}
\big[{\mathbb T}_{RR}^{(2,1)}\big]_{\alpha {\alpha'}}(x_\parallel,x'_\parallel) 
&=& \frac{1}{\lambda_R} \Big[\frac{1}{2} 
\delta_{\alpha {\alpha'}} (\partial_j \eta({\mathbf x}_\parallel))^2 - \delta_{\alpha i}
\delta_{\alpha' j} \partial_i \eta({\mathbf x}_\parallel) \partial_j
\eta({\mathbf x}'_\parallel) \Big] 
\delta^{(3)}(x_\parallel- x'_\parallel) \nonumber\\
\big[{\mathbb T}_{RR}^{(2,2)}\big]_{\alpha {\alpha'}}(x_\parallel,x'_\parallel) 
&=&\frac{1}{4} \big(\partial_\beta \partial'_\beta \delta_{\alpha\alpha'} - 
\partial_{\alpha'} \partial'_\alpha \big) 
 [ \eta({\mathbf x}_\parallel) - \eta({\mathbf x}'_\parallel) ]^2 
\int \frac{d^3k}{(2\pi)^3} e^{i k \cdot (x_\parallel - x'_\parallel)} \, |k| \nonumber\\
\big[{\mathbb T}_{RR}^{(2,3)}\big]_{\alpha {\alpha'}}(x_\parallel,x'_\parallel) 
&=& \epsilon^{i\alpha\beta} \epsilon^{j\alpha'\beta'}
\partial_i\eta({\mathbf x}_\parallel)
\partial'_j\eta({\mathbf x}'_\parallel)
\int \frac{d^3k}{(2\pi)^3} e^{i k \cdot (x_\parallel - x'_\parallel)} \,
\frac{k_\beta k_{\beta'}}{2 |k|} \nonumber\\ 
\big[{\mathbb T}_{RR}^{(2,4)}\big]_{\alpha {\alpha'}}(x_\parallel,x'_\parallel) 
&=&\frac{1}{4} \Big(\big[\partial_j \eta({\mathbf x}_\parallel)\big]^2
+
\big[\partial_j \eta({\mathbf x'}_\parallel)\big]^2 \Big) 
\int \frac{d^3k}{(2\pi)^3} e^{i k \cdot (x_\parallel - x'_\parallel)} 
\frac{k_\alpha k_{\alpha'}}{|k|}\;.
\end{eqnarray}

\subsection{Evaluation of $\Gamma^{(0)}$}
We recall that $\Gamma^{(0)}=\frac{1}{2} {\rm Tr}\ln\big[{\mathbb T}^{(0)}\big]$
where the trace runs over all the indices (Lorentz and indices that label
the two mirrors). 
To perform that trace it is convenient to note that 
\begin{eqnarray}
\widetilde{\mathbb T}^{(0)}(k) &=& \left( 
\begin{array}{cc} 
\frac{1}{\lambda_L} + \frac{|k|}{2} & \frac{|k|}{2} e^{- |k| a} \\
\frac{|k|}{2} e^{- |k| a} & \frac{1}{\lambda_R} + \frac{|k|}{2} 
 \end{array}\right) {\mathcal P}_\perp(k) \nonumber\\
&+& \left( 
\begin{array}{cc} 
\frac{1}{\lambda_L} + \frac{|k|}{2} & 0\\
0 & \frac{1}{\lambda_R} + \frac{|k|}{2} 
\end{array}\right) {\mathcal P}_\parallel(k) \;,
\end{eqnarray}
where we have introduced the transverse (${\mathcal P}_\perp$) and
longitudinal (${\mathcal P}_\parallel$) projectors, corresponding to the
$3$-vector $k$, namely, \mbox{$\big[{\mathcal P}_\perp\big]_{\alpha
\alpha'}(k) = \delta_{\alpha \alpha'} - \frac{k_\alpha k_{\alpha'}}{k^2}$},
and  \mbox{$\big[{\mathcal P}_\parallel\big]_{\alpha \alpha'}(k) =
\frac{k_\alpha k_{\alpha'}}{k^2}$}.
Since these projectors are orthogonal,
\begin{eqnarray}
\ln\big[{\mathbb T}^{(0)}\big] &=& {\mathcal P}_\perp(k) 
\ln 
\left(
\begin{array}{cc} 
\frac{1}{\lambda_L} + \frac{|k|}{2} & \frac{|k|}{2} e^{- |k| a} \\
\frac{|k|}{2} e^{- |k| a} & \frac{1}{\lambda_R} + \frac{|k|}{2} 
\end{array}
\right)  
\nonumber\\
&+& {\mathcal P}_\parallel(k)  \ln
\left(
\begin{array}{cc} 
\frac{1}{\lambda_L} + \frac{|k|}{2} & 0\\
0 & \frac{1}{\lambda_R} + \frac{|k|}{2} 
\end{array}
\right) \;, 
\end{eqnarray}
and
\begin{eqnarray}
{\rm Tr} \ln\big[{\mathbb T}^{(0)}\big] &=& 2 \times T L^2 \int \frac{d^3
k}{(2\pi)^3}
\ln \det 
\left(
\begin{array}{cc} 
\frac{1}{\lambda_L} + \frac{|k|}{2} & \frac{|k|}{2} e^{- |k| a} \\
\frac{|k|}{2} e^{- |k| a} & \frac{1}{\lambda_R} + \frac{|k|}{2} 
\end{array}
\right)  
\nonumber\\
&+& 1 \times T L^2\int \frac{d^3k}{(2\pi)^3} \ln \det 
\left(
\begin{array}{cc} 
\frac{1}{\lambda_L} + \frac{|k|}{2} & 0\\
0 & \frac{1}{\lambda_R} + \frac{|k|}{2} 
\end{array}
\right) \;. 
\end{eqnarray}

Discarding $a$-independent contributions, we see that $\Gamma^{(0)}$ may
then be written as follows:
\begin{eqnarray}\label{eq:gamma0result}
\Gamma^{(0)} \;&=&\; \frac{1}{2} \, L^2 \, T \, 2 \; \int \frac{d^3k}{(2\pi)^3} \,
\ln \Big[ 1 - \frac{ (\frac{|k|}{2})^2}{(\frac{1}{\lambda_L} +
\frac{|k|}{2})  (\frac{1}{\lambda_R} + \frac{|k|}{2})} e^{-2 |k| a} \Big]\nonumber\\
&\equiv&L^2\, T\, V_{\rm eff}(a) \;.
\end{eqnarray}
This result coincides with the vacuum energy corresponding to two imperfect, flat, and parallel mirrors
separated by a distance $a$, that we had computed  previously
(\cite{plb2008}) for the particular case $\lambda_L=\lambda_R$. Note that, given the boundary conditions
produced by our relativistic model (see Eqs.(\ref{eq:defssigma}) and (\ref{bc})), up to leading order the Casimir energy
for the electromagnetic field is twice the Casimir energy for the case of a scalar field. This was already shown in Ref. \cite{plb2008},
where it was also pointed out that for nonrelativistic matter the contributions of TE and TM modes are not equal, in agreement with the 
fact that the TE and TM reflection coefficients are different in this case \cite{barton2004}.

\subsection{Evaluation of $\Gamma^{(1)}$}
In the previous subsection we obtained the function $V_{\rm eff}$.
Although the evaluation of the term linear in $\eta$ is not necessary for
our next purpose of obtaining $Z$,  it is useful as an internal consistency
check of the calculations.

Recalling the expression for $\Gamma^{(1)}$, we see that we need the inverse of 
${\mathbb T}^{(0)}$. In Fourier space, it is given by:
\begin{eqnarray}\label{eq:t0inv}
\big[\widetilde{\mathbb T}^{(0)}(k)\big]^{-1} &=& \frac{1}{D(k)}\, 
\left( 
\begin{array}{cc} 
\frac{1}{\lambda_R} + \frac{|k|}{2} & - \frac{|k|}{2} e^{- |k| a} \\
-\frac{|k|}{2} e^{- |k| a} & \frac{1}{\lambda_L} + \frac{|k|}{2} 
 \end{array}\right) {\mathcal P}_\perp(k) \nonumber\\
&+& \left( 
\begin{array}{cc} 
\frac{1}{\frac{1}{\lambda_L} + \frac{|k|}{2}} & 0\\
0 & \frac{1}{\frac{1}{\lambda_R} + \frac{|k|}{2}}
\end{array}\right) {\mathcal P}_\parallel(k) \;,
\end{eqnarray}
where:
\begin{equation}\label{defD}
D(k) \,=\, \big(\frac{1}{\lambda_L} + \frac{|k|}{2}\big)
\big(\frac{1}{\lambda_R} + \frac{|k|}{2}\big)
-
\big( \frac{|k|}{2} e^{- |k| a} \big)^2 \;. 
\end{equation}

Then, using the notation: $\Delta_{ab} \equiv \big[{{\mathbb
T}^{(0)}}^{-1}\big]_{ab}$,
\begin{eqnarray}
\Gamma^{(1)} &=& \frac{1}{2} \, \Big[ \int_{x_\parallel, x'_\parallel}\;
 \big(\Delta_{LR}\big)_{\alpha {\alpha'}} (x,x')  
\;
\big({\mathbb T}^{(1)}_{RL}\big)_{{\alpha'}\alpha} (x',x) \nonumber\\
&+& \int_{x_\parallel, x'_\parallel}\; \big(\Delta_{RL}\big)_{\alpha {\alpha'}} (x,x')  
\;
\big({\mathbb T}^{(1)}_{LR}\big)_{{\alpha'}\alpha} (x',x) \Big] \;.
\end{eqnarray}
This may be evaluated explicitly by introducing Fourier transforms, the
result being:
\begin{equation}
\Gamma^{(1)} \;=\; \frac{T\,L^2}{2} \, \int \frac{d^3k}{(2\pi)^3} \,
\frac{|k|^3}{D(k)} \, e^{- 2 |k| a} \; \int d^2{\mathbf x}_\parallel \, \eta({\mathbf x}_\parallel)
\;.
\label{Gamma1fin}
\end{equation}
From Eqs.~(\ref{eq:gamma0result}) and (\ref{Gamma1fin}) one can easily show that
\begin{equation}
\Gamma^{(1)} \;=\; T\, L^2 V'_{\rm eff}(a) \int d^2{\mathbf x}_\parallel \, \eta({\mathbf x}_\parallel)
\;,
\end{equation}
as expected from Eq.(\ref{withlinear}).

\subsection{Evaluation of $\Gamma^{(2)}$}\label{sec:Gamma2}

We present the evaluation of the two contributions to $\Gamma^{(2)}$
separately. 

\subsubsection{ Contribution of $\Gamma^{(2,1)}$}\label{c123}
Let us consider first $\Gamma^{(2,1)}$.
It is convenient to recall the form of the inverse of ${\mathbb
T}^{(0)}$, presented in (\ref{eq:t0inv}), and of ${\mathbb T}^{(2)}$, in
equations (\ref{eq:tlr2}) and (\ref{eq:t2}). 

We first note, by explicit evaluation, that the terms 
$[{\mathbb T}_{LR}]^{(2)}$ and $[{\mathbb T}_{RL}]^{(2)}$ in
(\ref{eq:tlr2}) do not contribute to the force. 
We also see that ${\mathbb T}_{RR}^{(2,4)}$ in (\ref{eq:t2}) can be ignored, 
since its tensor structure allows it to mix only with the piece of 
$\big[{\mathbb T}^{(0)}\big]^{-1}$ which is proportional to ${\mathcal P}_\parallel$. 
Since neither object depends on $a$, the corresponding contribution is irrelevant to 
the calculation of Casimir forces.

On the other hand, only the `transverse', i.e. proportional to ${\mathcal
P}_\perp$, term in $\big[{\mathbb T}^{(0)}\big]^{-1}$ must be retained for
the rest of the terms. Indeed, it is the only part that can produce an
$a$-dependent contribution for the terms ${\mathbb T}_{RR}^{(2,1)}$ and
${\mathbb T}_{RR}^{(2,3)}$ in (\ref{eq:t2}).  On the other hand, the tensor
structure of ${\mathbb T}_{RR}^{(2,2)}$ allows it to mix only with the
transverse part of $\big[{\mathbb T}^{(0)}\big]^{-1}$. 

The results due the relevant terms, after extracting the term of second order in 
derivatives, shall have the following form:
 \begin{equation}
\frac{1}{2}{\rm Tr}\big[\Delta_{RR} {\mathbb T}_{RR}^{(2,b)}\big]
	\;=\; \frac{c_b(a)}{2}  \, T \int d^2{\mathbf x}_\parallel  \big[\partial_j \eta({\mathbf
	x}_\parallel)\big]^2 \;, 
\end{equation}
with $b = 1, 2, 3$.

Interestingly, ${T}_{RR}^{(2,1)}$ yields a local term (no need to perform a derivative
expansion):
\begin{equation}\label{c1}
	c_1(a) \,=\, \frac{1}{8}\int \frac{d^3k}{(2\pi)^3}  
	\frac{\vert{\mathbf k}_\parallel\vert^2}{ \lambda_R (\frac{1}{\lambda_L} + \frac{|k|}{2})
	(\frac{1}{\lambda_R} + \frac{|k|}{2})^2 
	\big[e^{2 |k| a} - \frac{(\frac{|k|}{2})^2}{(\frac{1}{\lambda_L} + \frac{|k|}{2})
	(\frac{1}{\lambda_R} + \frac{|k|}{2})}\big]} 
\end{equation}
(where we subtracted $a$-independent contributions).  We could replace
$\vert{\mathbf k}_\parallel\vert^2$ by $2/3\vert k\vert^2$ inside the above integral.
However, in its present form, Eq.(\ref{c1}) remains valid even in the case in which
$\lambda_R$ and $\lambda_L$ are functions of $k_0$.

For ${\mathbb T}_{RR}^{(2,2)}$,
\begin{eqnarray}
	c_2(a) &=& - \lim_{k \to 0} 
	\frac{\partial}{\partial \vert {\mathbf k}_\parallel\vert ^2} \int \frac{d^3p}{(2\pi)^3} |p+k|
	|p|^2  \frac{\frac{1}{\lambda_L} +
	\frac{|p|}{2}}{D(p)} \nonumber\\
	&=& -\frac{1}{8}
	\int \frac{d^3k}{(2\pi)^3}  
	\frac{  \left(1-\frac{1}{2}\frac{\vert{\mathbf k}_\parallel\vert^2}{k^2}\right)\vert k\vert^3}  
	{ (\frac{1}{\lambda_L} + \frac{|k|}{2})
	(\frac{1}{\lambda_R} + \frac{|k|}{2})^2 
	\big[e^{2 |k| a} - \frac{(\frac{|k|}{2})^2}{(\frac{1}{\lambda_L} + \frac{|k|}{2})
	(\frac{1}{\lambda_R} + \frac{|k|}{2})}\big]} \;,
\end{eqnarray}
where again we subtracted a term independent of $a$.

Finally, the contribution due to ${\mathbb T}_{RR}^{(2,3)}$ is also local,
and the result is $c_3(a)=-c_2(a)$.

\subsubsection{ Contribution of $\Gamma^{(2,2)}$}\label{c45}
The next point is the evaluation of $\Gamma^{(2,2)}$.
It can be shown, by using symmetries of the matrix elements appearing in
the expression, that this contribution reduces to:
\begin{equation}
\Gamma^{(2,2)} \;=\; G + U 
\end{equation}
where:
\begin{equation}
G \;=\; - \frac{1}{2} {\rm Tr} \big( \Delta_{LL} T^{(1)}_{LR} 
\Delta_{RR} T^{(1)}_{RL} \big)  \;,\;\;
U \;=\; - \frac{1}{2} {\rm Tr} \big( \Delta_{LR} T^{(1)}_{RL} 
\Delta_{LR} T^{(1)}_{RL} \big) \;. 
\end{equation}
After some algebra, we see that the result for $G$ may be
written as follows:
\begin{equation}
G = \frac{1}{2} \int \frac{d^3k}{(2\pi)^3}
|\widetilde{\eta}(k)|^2 {\mathcal G}(k) 
\end{equation}
\begin{equation}
{\mathcal G}(k) = - \frac{1}{4} \int \frac{d^3p}{(2\pi)^3}
e^{-  2 a |p + k|} \, \Big[\big(p \cdot (p + k) \big)^2 + p^2 (p + k)^2
\Big]
d_{LL}(p) d_{RR} (p + k) 
\label{gdek}
\end{equation}
where:
\begin{equation}
	d_{LL}(p) \equiv \frac{\frac{1}{\lambda_R} + \frac{|p|}{2}}{D(p)}  
	\;,\;\; d_{RR}(p) \equiv \frac{\frac{1}{\lambda_L} +
	\frac{|p|}{2}}{D(p)} \;.  
\end{equation}
Note that, as we are considering static surfaces, $\widetilde{\eta}(k)$ is proportional to
$\delta(k_0)$ and therefore ${\mathcal G}(k)={\mathcal G}(k_0=0, {\mathbf k}_\parallel)$
depends only on ${\mathbf k}_\parallel$.
Up to second order in derivatives, this term yields a contribution 
with the same form we had for $\Gamma^{(2,1)}$, this time with a coefficient:
\begin{equation}
c_4(a) =  \lim_{k \to 0} 
	\frac{\partial}{\partial \vert {\mathbf k}_\parallel\vert^2}{\mathcal G}(k)\, .
	\label{c4imp}
\end{equation}

On the other hand,
\begin{equation}
U = \frac{1}{2} \int \frac{d^3k}{(2\pi)^3}
|\widetilde{\eta}(k)|^2 {\mathcal U}(k) 
\end{equation}
\begin{equation}
{\mathcal U}(k) = - \frac{1}{4} \int \frac{d^3p}{(2\pi)^3}
e^{- (|p| + |p + k|) a} \, \Big[\big(p \cdot (p + k) \big)^2 + p^2 (p + k)^2
\Big]
d_{LR}(p) d_{LR} (p + k) 
\label{udek}\end{equation}
where:
\begin{equation}
d_{LR}(p) \equiv - \frac{|p|}{2}  \frac{e^{- |p| a}}{D(p)} \;.  
\end{equation}
Up to second order in derivatives, it produces  a coefficient:
\begin{equation}\label{c5imp}
c_5(a) =  \lim_{k \to 0} 
	\frac{\partial}{\partial\vert {\mathbf  k}_\parallel\vert^2}{\mathcal U}(k)\, .
\end{equation}
The explicit expressions for $c_4$ and $c_5$ can be obtained by computing
the derivatives in Eqs.~(\ref{c4imp}) and (\ref{c5imp}).

\subsection{The improved PFA in the electromagnetic case}

Using the results of the previous section, we can finally obtain our main result: the improved PFA for imperfect thin mirrors.
The zero-order contribution to the vacuum
energy is:
\begin{equation}\label{eq:e0result}
E_{\rm vac}^{(0)} \;=\; \int d^2{\mathbf x}_\parallel \, \int \frac{d^3k}{(2\pi)^3} \,
\ln \Big[ 1 - \frac{ (\frac{|k|}{2})^2}{(\frac{1}{\lambda_L} +
\frac{|k|}{2})  (\frac{1}{\lambda_R} + \frac{|k|}{2})} e^{-2 |k|
\psi({\mathbf x}_\parallel)} \Big]
\;.
\end{equation}
Putting together the results of section \ref{sec:Gamma2}, we see
that the NTLO correction to the PFA reads
\begin{equation}
E_{\rm vac}^{(2)} \,=\frac{1}{2} \int d^2{\mathbf x}_\parallel
\sum_{b=1}^{5}c_b(\psi)(\partial_j\psi)^2\, .
\end{equation}

These results can be immediately  generalized to the case in which
$\lambda_{L,R}$ become frequency dependent, for which the answer is
obtained by making the replacement $\lambda_{L,R}\rightarrow
\lambda_{L,R}(k_0)$ in the final expressions for the derivative expansion of the energy. This simple
generalization is valid because we are considering static surfaces: the
time-dependence of the modes of the vacuum field is trivial and therefore
one can treat independently  each frequency. By the same reason, the
generalization to the case in which the electromagnetic response of the
mirrors depends nontrivially on ${\mathbf k}_\parallel$ is not so trivial,
and would involve local momentum expansions at the different points of the
curved surface. This point deserves further investigation.

\section{Analysis of the results}

In this section we perform a general analysis of the results obtained.
Let us first consider the perfect-mirror limit.  One can derive the idealized limit of perfect conductivity 
by taking $\lambda_1, \lambda_2 \to \infty$. The leading term reads, in this case
\begin{equation}\label{eq:e0perf}
\Big[E_{\rm vac}^{(0)}\Big]_{\rm perf} \;=\; \int d^2{\mathbf x}_\parallel \, \int \frac{d^3k}{(2\pi)^3} \,
\ln \big[ 1 -  e^{-2 |k| \psi({\mathbf x}_\parallel)} \big] 
= -\frac{\pi^2}{720}\int \frac{d^2{\mathbf x}_\parallel}{\psi^3} \;,
\end{equation}
which is the well known PFA for perfect conductors.

The evaluation of the NTLO is more tedious. From the explicit expressions of 
the functions $c_1, c_2$ and $c_3$ presented in Section \ref{c123} we obtain:
\begin{eqnarray}
{c_1}_{\rm perf} &=& 0\nonumber\\
{c_2}_{\rm perf}&=&-{c_3}_{\rm perf}=-\frac{\zeta(3)}{12\pi^2\psi^3}\, .
\end{eqnarray}
It is worth to remark that, in this limit, ${c_1}_{\rm perf}+{c_2}_{\rm perf}+{c_3}_{\rm perf}=0$, that is, there is no contribution from $\Gamma^{(2,1)}$ for perfect conductors.

On the other hand, the computation of the functions $c_4$ and $c_5$ defined 
in Section \ref{c45} is less straightforward.  We write the integrals in
Eqs.~(\ref{gdek}) and (\ref{udek})
in spherical coordinates in momentum space, perform the derivatives with respect
to $\vert {\mathbf k}_\parallel\vert^2$, and finally compute the integrals. 
In this way we obtain
\begin{equation}
{c_4}_{\rm perf}={c_5}_{\rm perf}=\frac{15-\pi^2}{1080\psi^3}\, ,
\end{equation}
and therefore $Z(\psi)=\frac{15-\pi^2}{1080\psi^3}$, in agreement with the result obtained by Bimonte et al \cite{pfa_mit1}.

It is also of interest to analyze the opposite limit, in which the mirrors are almost transparent ( $\lambda_L,\lambda_R\ll \psi$).
The zeroth order vacuum energy can be easily obtained by expanding the
argument of the logarithm in Eq.~(\ref{eq:e0result}). The result is
\begin{equation}\label{eq:order0trans}
E_{\rm vac}^{(0)} \;=\; -\frac{1}{4}\int d^2{\mathbf x}_\parallel \, \int \frac{d^3k}{(2\pi)^3} \,\lambda_L\lambda_R
|k|^2   e^{-2 |k|\psi({\mathbf x}_\parallel)}
\;.
\end{equation}

The evaluation of the coefficients $c_b$ is also rather simple in this
limit. For instance, from Eq.~(\ref{c1}) we obtain
\begin{equation}\label{c1trans}
	c_1(\psi) \,=\, \frac{1}{8}\int \frac{d^3k}{(2\pi)^3}  
	\lambda_R \lambda_L |{\mathbf k}_\parallel|^2	
	 e^{-2 |k| \psi({\mathbf x}_\parallel)}\, .
	\end{equation}
As already mentioned, $c_2+c_3=0$. For almost transparent mirrors, the function  $D(p)$ defined in 
Eq.~(\ref{defD}) becomes $D(p)\approx (\lambda_L\lambda_R)^{-1}$. Therefore, up to quadratic order
in $\lambda_{L,R}$ we have 
\begin{equation}
	c_4(\psi) \,=\, -\frac{1}{4}\int \frac{d^3k}{(2\pi)^3}  
	\lambda_R \lambda_L    \left(\vert k\vert^2+\frac{1}{2}\vert{\mathbf k}_\parallel \vert^2\right)	
	 e^{-2 |k| \psi({\mathbf x}_\parallel)}\, ,
	\end{equation}
and $c_5=0$.  Combining these results we obtain
 \begin{equation}\label{eq:order2trans}
E_{\rm vac}^{(2)} \;=\; -\frac{1}{8}\int d^2{\mathbf x}_\parallel \, 
\big[\partial_j \eta({\mathbf
	x}_\parallel)\big]^2
	\int \frac{d^3k}{(2\pi)^3} \,\lambda_L\lambda_R
|k|^2   e^{-2 |k|\psi({\mathbf x}_\parallel)}
\;,
\end{equation}	
and the improved PFA is therefore
\begin{equation}\label{pfatrans}
E_{\rm vac} \;=\; -\frac{1}{4}\int d^2{\mathbf x}_\parallel \, \sqrt{g}
\int \frac{d^3k}{(2\pi)^3} \,\lambda_L\lambda_R
|k|^2   e^{-2 |k|\psi({\mathbf x}_\parallel)}
\;.
\end{equation}
Note that, in the semitransparent limit,  the NTLO correction  is equivalent to the insertion of the factor
$\sqrt{g}$, that is, the improved result corresponds to the use of the area of the curved surface in the usual PFA. It would be interesting to check if  this result is exact, as is the case for a scalar vacuum field
\cite{scalarexact}. It is also worth to note that 
the scaling of the energy with $\psi$ depends of course on the choice of $\lambda_{L,R}$: if both are constants, the integrand
in Eq.(\ref{pfatrans})  is proportional to $\psi^{-5}$. 

We end up this section with a discussion on the case of graphene-like
materials,  in which the electromagnetic properties of the mirrors are
described by dimensionless quantities.  As is well known, the charged
degrees of freedom in graphene can be effectively described by massless
fermions confined to the surface, with a propagation velocity $v_F\approx
1/300$. When the surface is flat, the interaction between charges and the
electromagnetic field is described by a vacuum polarization tensor which,
in our notation, would correspond to  a nonlocal interaction with
$\lambda(k)$ proportional to  $(k_0^2+{\mathbf k}_\parallel^2)^{-1/2}$,
after a rescaling of the temporal components of all vectors and tensors in
order to take into account that the fermions have a propagation velocity
different from $c=1$. For curved surfaces, we expect an effective
interaction of the form 
\begin{equation}\label{eq:graphene}
	{\mathcal S}_\Sigma (A; y) \;\simeq \;  \int
	d^3\sigma \, \sqrt{g(\sigma)}\, F_{\alpha\beta}\,  [\nabla^2]^{-1/2}F^{\alpha\beta} \;, 
\end{equation}
where $\nabla^2$ is the Laplacian on the surface (the above mentioned rescaling should also be applied to 
Eq.(\ref{eq:graphene})). The calculation of the Casimir energy for this particular case is of high interest, but is beyond the scope of the present paper. We will discuss here a toy model for graphene-like materials,
compatible with the assumptions we made so far. Therefore, we will consider $\lambda_{L,R}^{-1}=\xi_{L,R}\vert k_0\vert/2$,
where $\xi_{L,R}$ are dimensionless constants.  

By dimensional analysis, the improved PFA will have the same functional form than the case of perfect conductivity, that is
\begin{equation}
E_{\rm vac} \simeq \int d^2{\mathbf x_ \parallel} \;
\frac{1}{\psi^3}\left[\alpha_1(\xi_L,\xi_R)+\alpha_2(\xi_L,\xi_R)(\partial_j\psi)^2\right]\, .
\label{graph like}\end{equation}

\begin{figure}
\centering
\includegraphics[width=8cm , angle=0]{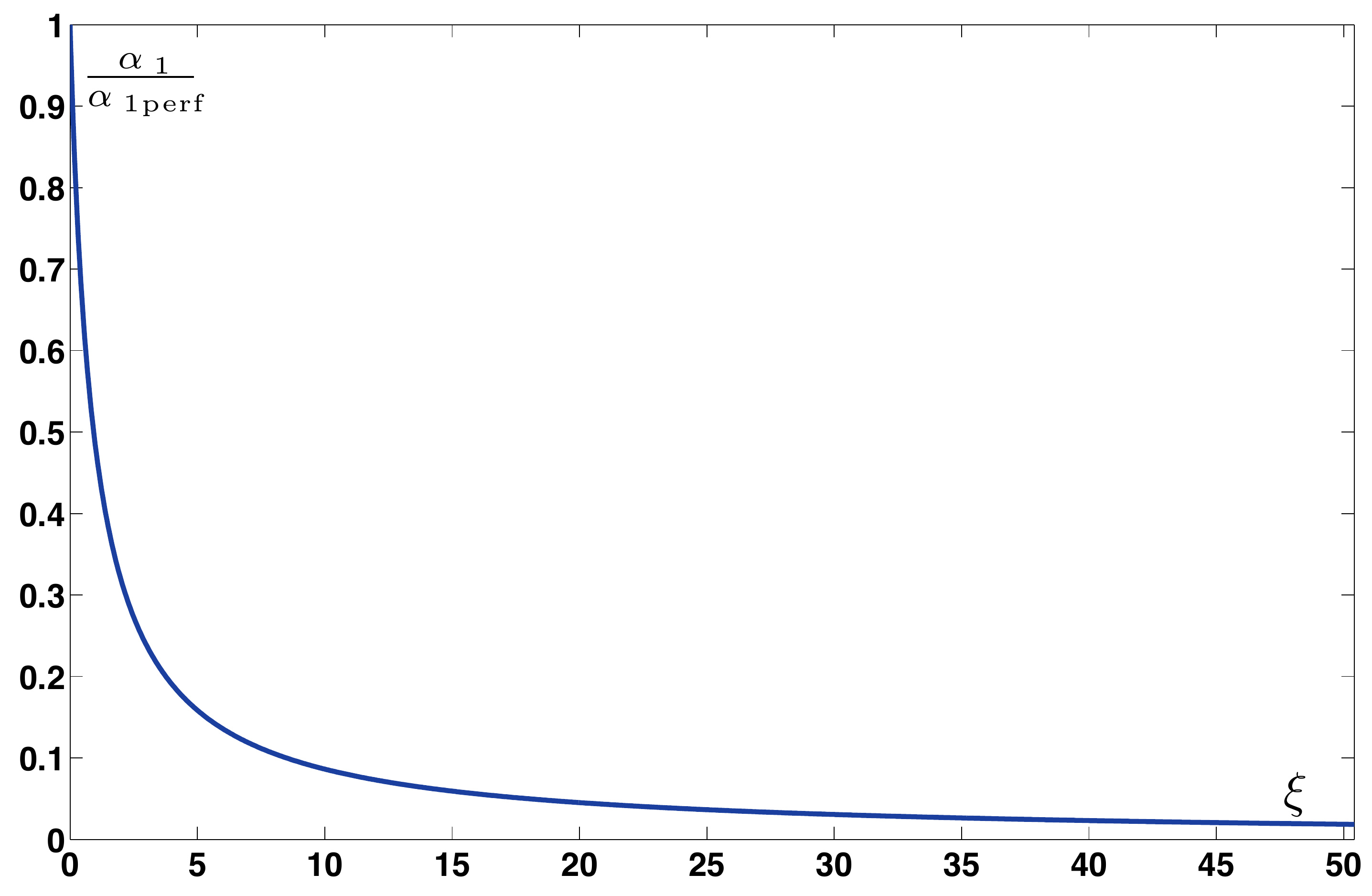}
\caption{Ratio between the coefficient of the leading term $\alpha_1(\xi)$ and the corresponding value for the perfect mirrors case, $\alpha_1(0)$,
as a function of the dimensionless parameter $\xi$. It is a monotonic decreasing function.} \label{E0}
\end{figure}

\begin{figure}
\centering
\includegraphics[width=8cm , angle=0]{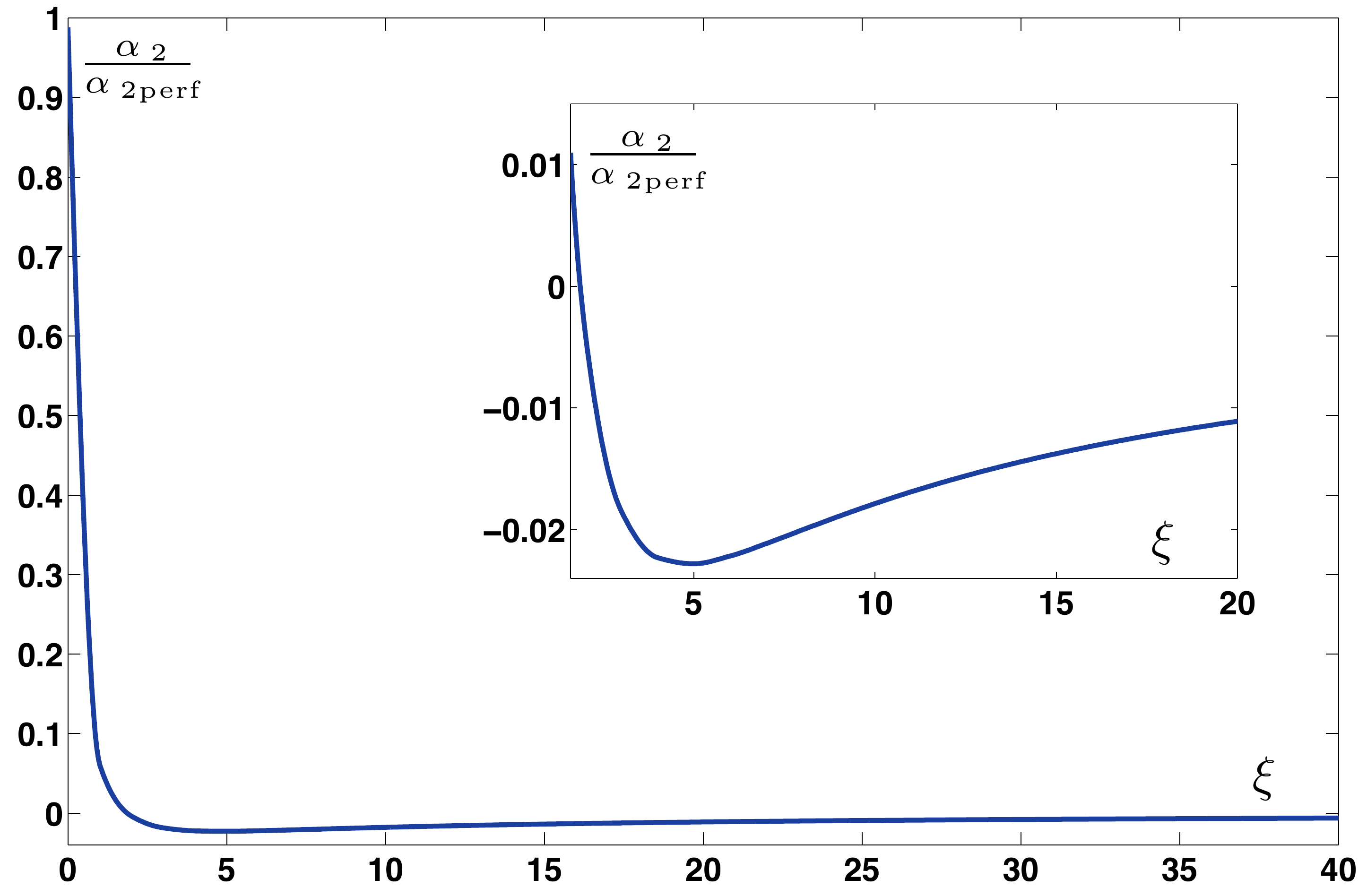}
\caption{Ratio between the coefficient of the NTLO correction $\alpha_2(\xi)$  and the corresponding value for the perfect case,
$\alpha_2(0)$,
as a function of the dimensionless $\xi$. The inset shows that this ratio is non-monotonous and changes sign for a particular value of $\xi$.} \label{cbs}
\end{figure}

The expressions for the dimensionless functions $\alpha_i$ can be easily
derived from our previous results.  We have computed numerically these
functions for the particular case $\xi_L=\xi_R=\xi$. The results are shown
in Figs. 1 and 2. As expected, these functions approach their
perfect conductivity limits for $\xi\to 0$ and  vanish in the almost transparent
limit, for $\xi\to \infty$. The leading order in the PFA has a rather simple
behavior: the absolute value of the coefficient $\alpha_1$ is a decreasing
function of $\xi$ (see Fig.1). The NTLO correction shows a qualitatively
different behavior, since it is non-monotonous and even changes sign at a
particular value of $\xi$ (Fig.2). Moreover, its absolute value falls faster
with $\xi$, that is,  for this kind of materials the NTLO correction
quickly loses relevance away from the infinite conductivity limit.

\section{Conclusions}

In this paper we have computed the Casimir energy for thin and imperfect
mirrors using a derivative expansion in the shape of the surfaces. The
leading term in the expansion reproduces the usual PFA, while the term
containing two derivatives represent the NTLO correction. These results
generalize previous works that involved perfect
mirrors~\cite{pfa_nos,pfa_mit1}, and may be regarded as complementary to
those for the interaction between thick mirrors~\cite{pfa_mit2}. 

The interaction between the mirrors and the vacuum field has been described
by a local effective action, which is a novel electromagnetic
generalization of the $\delta$-potentials usually considered for scalar fields.  
We also discussed some nonlocal generalizations, which 
could be useful to describe the interaction between curved graphene sheets.
To compute the vacuum energy we used a functional approach, and we used an
explicitly (electromagnetic) gauge invariant approach, whereby the
interaction term has been written in terms of vector auxiliary fields coupled to 
the Maxwell tensor. 

We have presented general expressions for the improved PFA for this model, and 
checked the particular limits corresponding to perfect conductors and almost-transparent mirrors.
For the particular case of mirrors described by a single dimensionless quantity $\xi$, we computed
the leading PFA and its NTLO correction as a function of $\xi$. We have found that the  NTLO
correction has a non-monotonous dependence on $\xi$, and that its absolute value drops quickly
for imperfect mirrors.

For the sake of simplicity, we considered a gently curved surface in front
of a plane. Moreover, we only considered the case in which the curved
surface can be described by a single function  $x_3=\psi(x_1,x_2)$. The
results can be extended to the case of  two curved surfaces described by
functions along the lines of Ref.\cite{pfa_mit1}.  The generalization to
cases in which the surfaces cannot be described in this way,
for example the case of an object inside another, is far from immediate. 

\section*{Acknowledgements}
This work was supported by ANPCyT, CONICET, UBA and UNCuyo.


\begin{thebibliography}{bib}
\bibitem{scalarexact} It has been shown in Ref.\cite{sqrtg} that in the almost transparent case the PFA gives the exact result for the {\it scalar} vacuum energy. We have checked that, when including the $\sqrt{g}$ in the interaction term,  the exact result  for the {\it scalar} case is the improved PFA (i.e. it is necessary to consider the terms containing two derivatives of $\psi$, and there are no corrections with higher derivatives up to $O(\lambda_s^2)$).   It is much more complicated to see whether this is true or not in the electromagnetic case for the model considered in this paper.
\end{thebibliography}
\end{document}